\documentclass[lettersize,journal]{IEEEtran}
\usepackage{amsmath,amsfonts}
\usepackage{array}
\usepackage[caption=false,font=normalsize,labelfont=sf,textfont=sf]{subfig}
\usepackage{textcomp}
\usepackage{stfloats}
\usepackage{url}
\usepackage{verbatim}
\usepackage{graphicx}
\usepackage{algorithmic}
\usepackage[linesnumbered,ruled,vlined]{algorithm2e}
\usepackage{tabularx}
\usepackage[square,numbers,sort&compress]{natbib}
\usepackage{xcolor}
\newcommand{\Input}{\textbf{Input:}} 
\newcommand{\Output}{\textbf{Output:}}
\begin{document}
\bstctlcite{IEEEexample:BSTcontrol}

\title{Cluster-Based Multi-Agent Task Scheduling for Space-Air-Ground Integrated Networks}
\author{Zhiying Wang, Gang Sun,~\IEEEmembership{Senior Member,~IEEE}, Yuhui Wang, Hongfang Yu,~\IEEEmembership{Member,~IEEE}, \\Dusit Niyato,~\IEEEmembership{Fellow, ~IEEE}
\thanks{This work was supported in part by the National Key Research and Development Program of China under Grant 2019YFB1802800.}
\thanks{Zhiying Wang, Gang Sun, Yuhui Wang, and Hongfang Yu are with School of Information and Communication Engineering, University of Electronic Science and Technology of China, Chengdu 611731, China (e-mail: zhiyingwang@std.uestc.edu.cn; gangsun@uestc.edu.cn; wyh410928@163.com; yuhf@uestc.edu.cn).}% <-this % stops a space
\thanks{Dusit Niyato is with the College of Computing and Data Science, Nanyang Technological University, Singapore (e-mail: dniyato@ntu.edu.sg).}
\thanks{The corresponding author: Gang Sun.}}
\maketitle

\begin{abstract}
The Space-Air-Ground Integrated Network (SAGIN) framework is a crucial foundation for future networks, where satellites and aerial nodes assist in computational task offloading. The low-altitude economy, leveraging the flexibility and multifunctionality of Unmanned Aerial Vehicles (UAVs) in SAGIN, holds significant potential for development in areas such as communication and sensing. However, effective coordination is needed to streamline information exchange and enable efficient system resource allocation. In this paper, we propose a Clustering-based Multi-agent Deep Deterministic Policy Gradient (CMADDPG) algorithm to address the multi-UAV cooperative task scheduling challenges in SAGIN. The CMADDPG algorithm leverages dynamic UAV clustering to partition UAVs into clusters, each managed by a Cluster Head (CH) UAV, facilitating a distributed-centralized control approach. Within each cluster, UAVs delegate offloading decisions to the CH UAV, reducing intra-cluster communication costs and decision conflicts, thereby enhancing task scheduling efficiency. Additionally, by employing a multi-agent reinforcement learning framework, the algorithm leverages the extensive coverage of satellites to achieve centralized training and distributed execution of multi-agent tasks, while maximizing overall system profit through optimized task offloading decision-making. Simulation results reveal that the CMADDPG algorithm effectively optimizes resource allocation, minimizes queue delays, maintains balanced load distribution, and surpasses existing methods by achieving at least a 25\% improvement in system profit, showcasing its robustness and adaptability across diverse scenarios.

\end{abstract}


\section{Introduction}

% 现有的发展，导致sagin的出现
% 随着iot，mec，ai的发展，无线网络服务激增
% 传统的地面网络难以支撑越来越多的网络需求
% 在海洋，沙漠，山区等地区地面网络往往没有好的覆盖
% sagin应运而生
\IEEEPARstart {T}{he} development of emerging technologies such as the Internet of Things (IoT), cloud computing, and artificial intelligence has led to a surge in wireless communication services. Nevertheless, traditional terrestrial networks encounter difficulties in satisfying the escalating traffic requirements and ensuring ubiquitous coverage. In some isolated and elevated areas, such as oceans and mountains, ground cellular or fiber networks suffer from inadequate or absent access due to topographical impediments \cite{shang2021computing}. After decades of advancement, the global space infrastructure has attained a high level of maturity and dependability, establishing a comprehensive space-based information system \cite{9712216}. By utilizing the space, aerial, and ground segments with modern information network technologies, the Space-Air-Ground Integrated Network (SAGIN) has emerged as a potential solution to overcome the shortcomings of ground communication systems \cite{liu2018space}. By achieving device information fusion and collaborative cooperation through Airborne Computing (AirComp), the low-altitude economy will provide critical support for applications such as network coverage, environmental monitoring, disaster relief, and logistics delivery \cite{jiang20246gnonterrestrialnetworksenabled}.

% sagin是一个融合地面，空中，卫星，海洋的基础框架，可以解决单一地面网络情况的局限性，实现对地球的广泛覆盖。
% aerial-based网络包括hap和uav网络，可以给地面网络提供支持，为用户提供灵活高效的服务
% space-based网络由多颗卫星组成，这些卫星将形成一个多连接多层次的数据传输和信息处理系统，并且与aerial-based网络协同，对地面实现更广阔的网络覆盖。

SAGIN is a foundational framework that integrates terrestrial, aerial, satellite, and maritime networks, addressing the limitations of standalone terrestrial networks to enable broad coverage across the globe. The aerial-based network component, which includes High-Altitude Platforms (HAPs) and Unmanned Aerial Vehicles (UAVs), offers flexible and efficient support to ground networks \cite{kandeepan2014aerial}. This aerial layer can dynamically allocate resources based on user demand and adjust its coverage area as needed, making it ideal for disaster response, remote sensing, and rural connectivity. The space-based network layer, consisting of multiple satellites, forms a multi-link, multi-tiered data transmission, and information processing system \cite{liu2020development}. This integrated network structure supports applications such as real-time data collection for climate monitoring, maritime navigation, and seamless broadband service for remote communities.

With the rapid development of devices like smartphones and tablets, a variety of computation-intensive and latency-sensitive applications have emerged. Mobile devices, however, often have limited processing capabilities. To address this, task offloading has been proposed, allowing user devices to offload computational tasks to nearby high-performance computing centers \cite{tang2021mobile,qiu2022mobile}. This offloading not only reduces computational delay but also conserves the energy consumption of mobile devices. However, the complex network architecture introduces numerous challenges for resource scheduling and management. These include issues such as cooperative control and collaborative data transmission in heterogeneous networks, as well as mobility management in time-varying networks \cite{10571777,10113881}. Thus, efficiently scheduling tasks has become a focal point for researchers, who are actively exploring methods to optimize resource allocation, improve network responsiveness, and enhance the overall quality of service in mobile edge computing environments \cite{10234628}.

There is a substantial body of research on task scheduling in various SAGIN scenarios. The authors in \cite{9810267} addressed the task scheduling issue in latency-aware IoT networks and proposed a non-preemptive priority queuing and cross-layer optimization scheme. In \cite{10297374}, the authors introduced a Medium Access Control (MAC) protocol for controlling massive IoT device access in a network consisting of a satellite, a high-altitude platform (HAP), and multiple IoT device networks, providing differentiated services to IoT devices with varying Quality of Service (QoS) requirements. The authors in \cite{10714036} proposed an intelligent offloading decision algorithm for resource competition scenarios based on Proportional Fairness-Aware Auction and Proximal Policy Optimization (PPO). Finally, the authors in \cite{9222519} formulated the online scheduling problem as an energy-constrained Markov Decision Process (MDP) and proposed a deep risk-sensitive reinforcement learning algorithm.

However, existing centralized control schemes typically focus only on the strategy selection of individual decision-makers, overlooking the complex impacts of information exchange and parameter sharing between them. At the same time, fully distributed schemes lack coordination and face significant convergence challenges, making it difficult to achieve optimal results for the overall system. To address this issue, this paper proposes a Clustering-based Multi-Agent Deep Deterministic Policy Gradient (CMADDPG) method in SAGIN, which adopts a cooperation mode of centralized training and distributed execution, effectively overcoming these challenges. The main contributions of this paper are as follows:

\begin{itemize}
    \item To reduce communication overhead among nodes in large-scale UAV scenarios and adapt to dynamic topology changes, we propose a low-complexity K-Means-based dynamic UAV clustering algorithm (KMDUC), which includes a centralized clustering phase and a distributed cluster maintenance phase, thereby achieving a hybrid control mode of distributed and centralized control.
    \item Next, we propose a cooperative offloading algorithm based on the multi-agent deep deterministic policy gradient, which employs a centralized training and distributed execution framework. During the training phase, a centralized critic network deployed on the satellite performs joint training. In the execution phase, each agent deployed on a UAV only uses local observation information to make decisions through its distributed actor network, achieving both global optimization and locally efficient decision-making for large-scale SAGIN.
    \item Numerical simulation experiments validate the effectiveness of the proposed algorithm. The results show that the proposed algorithm demonstrates advantages in system profit, load balancing, and fairness, particularly surpassing existing methods by achieving at least a 25\% improvement in system profit, which comprehensively considers task data size, computational complexity, and delay requirements.
\end{itemize}

The remainder of this paper is organized as follows. Section II presents the related work. We describe the system models and problem formulation in Section III. In Section IV, we give the details of our proposed algorithm. Section V presents the simulation results, followed by the conclusion in Section VI.

\section{Related Work}

With the development and widespread adoption of 5G networks, there has been an increase in computationally intensive and latency-sensitive tasks. To meet user demands and support the evolution of future 6G networks, a novel SAGIN architecture has emerged \cite{9722775}. SAGIN primarily comprises three segments: space, air, and ground, which cooperate to achieve global communication coverage and provide uninterrupted network services. However, this integration also faces numerous challenges: i) real-time responsiveness; ii) integration complexity; and iii) scalability.
% 1.SAGIN 任务调度概述
% 1.1框架概述
\subsection{Resource Allocation In SAGIN}
% 1.2任务调度概述
Due to the large number of nodes and links in SAGIN, the network complexity significantly exceeds that of conventional networks. Moreover, the channel and computational resources within the network are limited, making efficient resource scheduling essential to maximize resource utilization and achieve better service performance. In \cite{9749937}, the authors proposed a service function chain mapping method based on delay prediction, calculating the deployment delay for each path and selecting the one with the minimum delay for chain deployment, effectively improving the Central Processing Unit (CPU) and link resource utilization. In \cite{10454605}, a Lyapunov algorithm was employed to ensure queue stability while maximizing long-term network utility, maintaining throughput and fairness among IoT nodes. In \cite{10001422}, the authors considered a network comprising satellites and UAVs, aiming to minimize the weighted energy consumption of ground users (GUs) and UAVs while meeting maximum delay constraints. Additionally, \cite{10494938} analyzed the use of mathematical optimization, game theory, and AI techniques (e.g., deep learning and reinforcement learning) to minimize latency and maximize network throughput and stability, contributing to the advancement of next-generation SAGIN networks. Although many studies have explored resource scheduling, designing an algorithm for SAGIN with numerous nodes remains challenging. It is crucial to balance node cooperation and resource fairness, which has become a key focus of research for optimizing network performance.

\subsection{Traditional Optimization Algorithms}
% 2.一般优化模型（合作）
Many researchers have proposed traditional optimization algorithms to address resource scheduling challenges in SAGIN scenarios. The authors in \cite{10008603} utilized Successive Convex Approximation (SCA) to solve a non-convex problem by iteratively optimizing user association, offloading decisions, and UAV positioning, aiming to minimize the weighted energy consumption while satisfying maximum delay constraints. The work in \cite{10304325} introduced the Slice-Soft-SAGIN framework, which allocates slicing resources by leveraging wireless spectrum and computational resources to ensure each slice’s quality of service, achieving efficient resource allocation. To maximize the weighted sum rate, the authors in \cite{he2024symbioticsagininteroperatorresource} proposed both a centralized algorithm based on SCA and a distributed algorithm based on the Alternating Direction Method of Multipliers (ADMM), with both approaches validated for effectiveness. Furthermore, the authors in \cite{9874801} tackled the NP-complete three-way matching problem in resource allocation by introducing constraints to transform it into a restricted three-way matching problem, facilitating optimization. While traditional optimization algorithms have shown effectiveness in specific resource scheduling problems in SAGIN, they often face limitations in scalability, adaptability, and computational efficiency, especially in dynamic and complex environments.

\subsection{Deep Reinforcement Learning (DRL) Algorithms}
Traditional optimization algorithms are often constrained by problem convexity and high complexity in large-scale scenarios, leading many researchers to adopt DRL for its scalability and flexibility.
% 3.强化学习模型
% 3.1单智能体
\subsubsection{Single-Agent DRL}
In \cite{10398221}, the authors first established a SAGIN architecture based on Software-Defined Networking (SDN) and Network Function Virtualization (NFV) and designed a resource scheduling and computation offloading algorithm using Deep Q-Network (DQN). In \cite{10118888}, the Asynchronous Advantage Actor-Critic (A3C) algorithm was employed to effectively reduce the action space and accelerate convergence in SAGIN-assisted vehicular networks, resulting in improved rewards. The study in \cite{huang2024fairresourceallocationhierarchical} used satellites with Inter-Satellite Links (ISL) as cloud servers and designed UAV trajectories, user-UAV associations, and UAV-satellite associations to complete federated learning tasks within a limited service time while ensuring service fairness.

% 3.2多智能体（合作场景）
\subsubsection{Multi-Agent DRL}
As the scale of SAGIN expands, Multi-Agent DRL can enhance overall performance through communication and collaboration among agents. In \cite{10158321}, the authors designed a Multi-Agent DRL algorithm based on the Actor-Critic framework. Compared to single-agent reinforcement learning, the Critic network in this framework can simultaneously evaluate the actions produced by the Actor network and coordinate the actions of multiple actors to improve reward. To address multi-objective optimization for throughput, delay, and slice coverage, the authors in \cite{10560515} decomposed the problem into two sub-problems and solved it using the MADDPG algorithm. In \cite{10502326}, the traffic offloading problem in SAGIN was modeled as a Decentralized Partially Observable Markov Decision Process (DEC-POMDP) and solved with the Differentiated Federated Soft Actor-Critic (DFSAC) algorithm. Additionally, the authors proposed a Quantum Multi-Agent Reinforcement Learning (QMARL) approach, which improves global access availability and energy efficiency in large-scale SAGIN by reducing the action space dimensionality.

Single-agent approaches depend on centralized control, which limits scalability and flexibility in large, dynamic environments. While multi-agent systems improve scalability and flexibility, they face challenges such as communication overhead and coordination issues, leading to slower convergence and reduced stability. To overcome these limitations, we propose a novel algorithm that combines the advantages of both centralized and distributed approaches, ensuring scalability, flexibility, and efficient coordination in dynamic SAGIN environments.

\section{System Model}
 
\begin{table}[ht]
\centering
\caption{Explanation of Variables Used in This Paper}
\label{table1}
\begin{tabular}{ll}
\hline
\textbf{Variable} & \textbf{Description} \\ \hline
$\alpha_{k,j}$          & Offloading decision for task $k$ to device $j$. \\ \hline
$\mathcal{B}$           & Set of base stations. \\ \hline
$c_n^*$                 & Optimal number of clusters. \\ \hline
$C_j(t)$                & Computational capacity of target device $j$ at time $t$. \\ \hline
$d^c_{k,j}$             & Computing delay for task $k$ on device $j$. \\ \hline
$d^{tran}_{k,j}$        & Transmission delay for offloading task $k$ to device $j$. \\ \hline
$\delta$                & Allowable delay. \\ \hline
$E[N_l]$                & Expected number of UAVs in cluster $l$. \\ \hline
$f_j$                   & Computing capacity of device $j$. \\ \hline
$G$                     & Task profit function. \\ \hline
$J$                     & Objective function for position-based K-Means clustering. \\ \hline
$\mathcal{J}$           & Set of target computing devices. \\ \hline
$\kappa$                & Density of UAVs per unit area. \\ \hline
$\lambda$               & Delay sensitivity parameter. \\ \hline
$loc$                   & Position of the UAV. \\ \hline
$\mathbf{M}$            & Motion model of UAVs. \\ \hline
$N_{\text{BS}}$         & Total number of base stations. \\ \hline
$N_u$                   & Total number of UAVs. \\ \hline
$P_{il}$                & Coverage probability of UAV $i$ under cluster head $l$. \\ \hline
$P_{max}$               & Maximum coverage probability for a UAV. \\ \hline
$\phi$                  & Data size (in bits). \\ \hline
$\rho$                  & Computational workload (CPU cycles per bit). \\ \hline
$R$                     & Communication radius of the UAV. \\ \hline
$R_{ij}$                & Data rate between UAV $i$ and target device $j$. \\ \hline
$\mathbf{R^*}$          & Computational resource vector within cluster coverage. \\ \hline
$s$                     & Index of the Low-Earth Orbit (LEO) satellite. \\ \hline
$T_{cls}$               & Clustering period for the UAV network. \\ \hline
$t$                     & Time unit for discrete time slots. \\ \hline
$\tau_{\text{prop}}$    & Propagation delay based on communication distance. \\ \hline
$\mathcal{T}$           & The set of all time slots. \\ \hline
$\mathcal{U}$           & Set of UAVs. \\ \hline
$U_i$                   & UAV $i$. \\ \hline
$v$                     & Speed of the UAV. \\ \hline
$\eta_k$                & Computation priority of task $k$ on the target device. \\ \hline
$\theta$                & Change in the direction angle of the UAV's motion. \\ \hline
\end{tabular}
\end{table}

\subsection{The SAGIN Architecture}

The SAGIN comprises three interconnected layers: the space-based network, the air-based network, and the ground-based network. Ground stations provide high computational capabilities, helping UAVs offload computational tasks. However, because ground stations are fixed and immobile with relatively low altitudes, their coverage area is limited. In contrast, the LEO satellite constellation operates at higher altitudes, offering continuous coverage, but it introduces additional propagation delays in the UAV-satellite link. Therefore, how to leverage the distinct characteristics of UAVs, ground stations, and the LEO satellite constellation, and implement effective task scheduling strategies to allocate computational tasks appropriately to different destination network components in the SAGIN, aiming to maximize profits and task completion rates, becomes the key focus of this paper.

\begin{figure}[!t]
	\centering
	\includegraphics[width=0.48\textwidth]{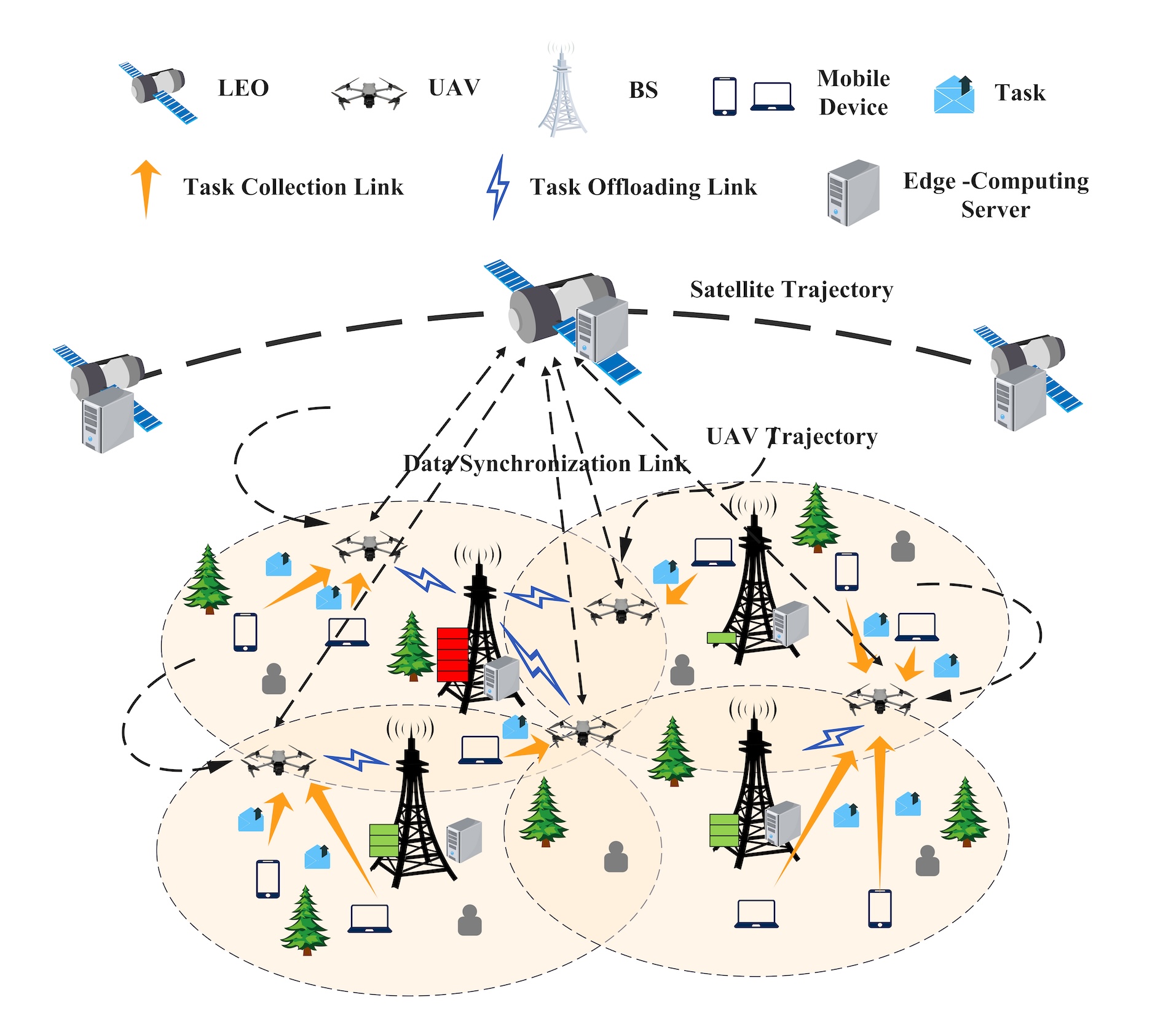}
	\caption{Task scheduling cooperation under SAGIN.}
	\label{sagin}
\end{figure}

As shown in Figure \ref{sagin}, UAVs are represented by the set \( \mathcal{U} = \{ U_i \mid i = 1, 2, \dots, N_u \} \), where \( U_i \) represents UAV \( i \), and \( N_u \) is the total number of UAVs. UAVs often belong to the same service provider or collaborate across providers to optimize the utilization of network computing resources during task offloading. By sharing information about factors such as distance, network load, and task requirements, UAVs can coordinate their offloading decisions to balance the load across multiple base stations (BSs). This cooperative approach helps to minimize task backlog and congestion at any single BS, reducing queuing delays and improving task completion rates. Additionally, this strategy ensures a more efficient use of available computing resources, achieving better load balance and enhancing overall system performance.

In our network scenario, we adopt a specific mobility model, denoted as $\mathbf{M} = (loc, \theta, v)$, where ground stations are stationary with fixed locations. The time duration of each time slot is denoted as $t$ and $\mathcal{T}$ is the set of all time slots. The UAV’s speed $v$ remains constant for a time slot, and its direction changes randomly within a certain angle $\theta$ at each time slot.

In this paper, we adopt a discrete timeslot-based approach. At each timeslot, the UAV's position is denoted as $loc_t$. As the UAV moves and its position changes, the set of available target network nodes $\mathcal{D}_t$ for task offloading also changes. During each timeslot, the UAV can choose multiple computational tasks to offload to appropriate ground stations or LEO satellites. However, we adopt a 0-1 offloading strategy \cite{8672604}, meaning that each task can only be offloaded to one target node for computation.

After the UAV collects tasks from mobile terminals if these tasks have not been processed or offloaded to a ground station or LEO satellite, they are stored in the UAV's task queue. Subsequently, the UAV can process or offload the computational tasks from the task queue to other nodes. Therefore, the UAV needs to reasonably schedule the processing order of tasks in the queue to achieve orderly and intelligent scheduling.

\subsection{Task Model}
We define a tuple $\mathbf{Task_k} = (\phi, \rho, \delta, G)$ to describe computing task $k$. Here, $\phi$ represents the input data size in bits, and $\rho$ indicates the computational workload in terms of CPU cycles per bit, i.e., the number of CPU cycles required to process one bit of data. Additionally, $\delta$ denotes the allowable delay for the task, with only those tasks completed within the deadline yielding revenue. The task profit function \cite{lin2020survey} is expressed as:
\begin{equation}
    G = \phi \cdot \rho \cdot e^{-\lambda \cdot \delta},
\end{equation}
which reflects the reward for successfully completing the task. The parameter $\lambda$ adjusts the sensitivity to delay, with a higher $\lambda$ emphasizing stricter delay requirements. The exponential factor $e^{-\lambda \cdot \delta}$ acts as a discount, reducing the task profit as the tolerable delay increases. Thus, the task profit depends on the data size, computational load, and delay tolerance.

\subsection{Computing Model}

Given the limited computational capacity of UAVs, tasks can be offloaded to satellites, BSs, or processed locally. We consider a satellite \( s \), which is primarily responsible for providing coverage, task transmission, and computation. There are \( N_{BS} \) base stations, and the set of base stations is denoted as \( \mathcal{B} = \{ 1, 2, \dots, N_{BS} \} \). Therefore, the task can be processed at a destination device $j \in \mathcal{J} = \{ 0 \cup \mathcal{B} \cup s \}$, where \( j = 0 \) means the task is processed locally, \( 1 \leq j \leq N_{BS} \) means the task is offloaded to BS, and $j = s$ means the task is offloaded to the satellite for computation.

The tuple $(\alpha_{k,j}, \eta_k)$ represents the offloading decision for task $k$. $\alpha_{k,j}$ denotes the 0-1 offloading decision. When task $k$ is offloaded to device $j$, $\alpha_{k,j} = 1$; otherwise, $\alpha_{k,j} = 0$. $\eta_k$ indicates the computation priority of the task on the target device. The second part of the tuple, $\eta_k \in [0,1]$, represents the relative weight or bias assigned to the computation priority of task $k$ on the selected target device. This reflects the computational order of task $k$, which prioritizes tasks with tighter delay constraints or deadlines. This variable allows for adaptive priority assignment, ensuring that delay-sensitive tasks can receive preferential treatment. Based on the computation priority, the target device processes the tasks sequentially in accordance with the assigned computational queue.

In addition, to avoid conflicts or congestion caused by the simultaneous transmission of multiple tasks, we assume that new tasks cannot be forwarded until the offloading process of the preceding task is completed. We define the computing capacity of BS as $f_j$ and the computing capacity of the satellite as $f_0$, both measured in CPU cycles per second. The delay in offloading task $k$ consists of two parts: queuing delay and computing delay. The computing delay for all tasks at the offloading destination device $j$ can be determined using the following formula:
\begin{equation}
	\label{d_compu}
	d_{k,j}^{c} = \frac{{\alpha_{k,j} \cdot \phi_k}}{{\rho_k \cdot f_j}},
\end{equation}
where $\phi_k$ represents the total computational demand of task $k$, measured in CPU cycles. $\rho_k$ denotes the parallelization efficiency coefficient of task $k$, with a range of $0 < \rho_k \leq 1$. It reflects the task's capability for parallel processing. The closer the value is to 1, the easier it is for the task to achieve efficient parallelism.

The queuing delay of a computing task on the target device is determined by the cumulative computing delays of all preceding tasks in the execution order. However, as it is influenced by other scheduling decisions, its exact mathematical expression cannot be defined.

\subsection{Communication Model}

In this paper, we assume that BSs and LEO satellites operate on different frequency bands, significantly reducing communication interference between the two \cite{wang2020multi}. Task offloading is achieved via wireless communication, with factors such as communication distance and rain attenuation impacting the quality of the wireless link. Assuming constant weather conditions during the task offloading process, the channel gain of the communication link between the UAV and the LEO satellite or BS can be approximated as dependent on the UAV's location. We employ a two-ray path loss model that accounts for both the direct path and the ground reflection path. The path loss (PL) between the UAV and the target device \cite{tang2021deep} is expressed as:
\begin{equation}
	PL_{ij} = 20\log_{10}\left(\frac{4\pi d_0}{\lambda}\right) + 10n\log_{10}\left(\frac{d}{d_0}\right) + X_\sigma,
\end{equation}
where, $d$ represents the actual communication distance between the UAV and the target device, while $d_0$ denotes the reference distance. $\lambda$ indicates the wavelength of the electromagnetic wave, defined as $\lambda = \frac{c}{f}$, where $c$ is the speed of light, and $f$ is the signal frequency. $n$ is the path loss exponent, determined by the surrounding propagation environment. $X_\sigma$ is a Gaussian random variable. These parameters will depend on the specific characteristics of the communication system in use. Consequently, the data rate $R_{ij}(t)$ for the link between the UAV and the target device can be calculated as follows:

\begin{equation}
	R_{ij} = B \cdot \log_{2}\left(1 + \frac{P_T \cdot G_T \cdot G_R}{P_N \cdot PL_{ij}(d)}\right),
\end{equation}
where, $B$ represents the bandwidth, $P_T$ is the transmit power, $G_T$ is the transmit antenna gain, $G_R$ is the receive antenna gain, $P_N$ is the noise power, and $PL_{ij}(d)$ is the path loss between the UAV and the satellite.

Let $d_{k,j}^{tran}$ represent the transmission delay for offloading a task to destination $j$, which can be expressed as:

\begin{equation}
	\label{d_tran}
	d_{k,j} ^{tran}= \frac{\alpha_{k,j}{\phi_k}}{R_{ij}} + \tau_{prop}.
\end{equation}

The propagation delay $\tau_{prop}$ depends on the value of $j$. When the communication link is between the UAV and a LEO satellite, i.e., $j = 0$, the propagation delay is calculated as the distance between the two divided by the speed of light. However, when the communication link is between the UAV and a BS, i.e., $j \neq 0$, the propagation delay is ignored due to the shorter communication distance. Therefore, $\tau_{prop}$ is expressed as:

\begin{equation}
	\tau_{prop} = 
	\begin{cases}
		\frac{d_{k,j}}{c}, & \text{if} \quad j = s \\
		0, & \text{otherwise}.
	\end{cases}
\end{equation}

The total delay of the task includes processing delay and transmission delay.

\begin{equation}
	d_{k,j} ^{total} = d_{k,j} ^{tran}+d_{k,j} ^{c}
\end{equation}

\subsection{Problem Formulation}

In this paper, we aim to maximize the long-term profits of the system in cooperative scenarios while satisfying the task delay constraint. Therefore, the task scheduling problem of SAGIN is formulated as:

\begin{align}
\label{problem}
	\max_{\boldsymbol{\alpha}_{k,j}}   &\lim_{T \to \infty} \frac{1}{T}\sum_{t\in \mathcal{T}} \sum_{j\in \mathcal{J}} \sum_{k\in \mathcal{L}} \alpha_{k,j}(t) G_{k}(t) \\
	\text{s.t.}
	\label{C1}
	& \quad \sum_{j=1}^{M} \alpha_{k,j} = 1, \quad \forall k \\
	\label{C2}
	& \quad d_{k,j} \leq \delta_{k}, \quad \forall i,j,k \\
	\label{C3}
	& \quad \alpha_{k,j} \in \{0,1\}, \quad \forall j,k \\
	\label{C4}
	& \quad \sum_{i=1}^{N} \sum_{k=1}^{L} \alpha_{k,j}(t) \phi_{k} \gamma_{k}  \leq C_{j}(t), \quad \forall j,t.
\end{align}

We seek to maximize the system's long-term average profit. The objective function is the limit of the average profit as the time horizon tends to infinity. The average profit is calculated by summing the product of the offloading decision $\alpha_{k,j}(t)$ and the task profit $G_{k}(t)$ over all time steps $t$, offloading destinations device $j$, and task $k$.

The constraints of the problem are as follows: Equation (\ref{C1}) indicates that each task can only be offloaded to one destination, meaning that, collectively, for each task $k$, the sum of the offloading decisions to all destinations should equal 1; Equation (\ref{C2}) ensures that the total delay for each task at each offloading destination does not exceed its specified deadline; Equation (\ref{C3}) indicates that the offloading decision $\alpha_{k,j}$ is binary, taking values 0 or 1, representing whether the task is offloaded to a specific destination; Equation (\ref{C4}) is a constraint on the total offloading load for each offloading target $j$, where $\phi_{k}$ is the input data size, $\gamma_{k}$ is the computational workload, and $C_{j}(t)$ is the computational capacity of the offloading target, ensuring that the computational resources of each offloading target device are not overloaded.

The optimization problem outlined in Equation (\ref{problem}) is an integer nonlinear optimization problem, complicated by an unknown number of newly collected tasks in each epoch. The evolving ergodic nature of the UAV’s location, along with the task queue backlog and the interference arising from UAV offloading decisions, classifies the problem as NP-hard \cite{9475463}.

\section{Algorithm Design}

This paper addresses the challenge of cooperative task scheduling in an integrated space-air-ground network, with the goal of maximizing system profit, as formulated in problem (\ref{problem}). A key challenge is designing a coordination mechanism that minimizes both algorithmic complexity and system overhead. While centralized control can maximize profit, it incurs high signaling costs, whereas fully distributed schemes suffer from a lack of coordination. To overcome these challenges, we propose the CMADDPG algorithm. This approach combines dynamic clustering and cooperative offloading strategies to enable efficient task scheduling, coordinated control, and enhanced system profit within the network.

\subsection{Dynamic UAV Clustering Algorithm}

The high-speed movement and dynamic nature of UAV networks lead to frequent topology changes, increasing network management and resource scheduling costs \cite{gupta2015survey}. UAV clustering is an effective method for improving performance in large-scale UAV networks by dividing the network into clusters, each with a Cluster Head (CH) for communication. However, traditional clustering schemes often neglect UAV mobility and relative positioning, failing to adapt to the dynamic nature of self-organizing networks. Frequent cluster updates increase control overhead and delays. To address this, this paper proposes a KMDUC, which considers UAV mobility while maintaining clustering stability.

The clustering algorithm divides the UAV network into several clusters, each consisting of a CH and several Cluster Members (CMs). In this scenario, the CH is responsible for communicating with other CHs or satellites and issuing task offloading decisions to CMs, while CMs are responsible for collecting computing tasks and communicating with the CH.

In summary, the KMDUC algorithm consists of two stages: the clustering stage and the cluster maintenance stage. In the clustering stage, a preliminary clustering scheme is determined using a centralized method based on UAV location information, with satellites collecting data over a period \( T_{\text{cls}} \), calculating clustering results, and broadcasting them to the UAVs. During the cluster maintenance stage, the UAVs' locations change dynamically, requiring a distributed maintenance algorithm to adjust the clusters in real time, ensuring their stability. The centralized method in the clustering stage provides a global view for better clustering results, while the distributed algorithm in the maintenance stage offers flexibility, lower control overhead, and real-time adaptability. Together, these stages enable KMDUC to leverage cooperative information and dynamically adjust to the high mobility and self-organizing nature of the integrated space-air-ground network.

\subsubsection{Position-based Clustering Algorithm}
For the clustering phase, the main factors affecting the efficiency of UAV task scheduling include computational delay, queuing delay, and transmission delay, all of which are closely related to the position of the UAVs. In the scenario considered in this paper, since each UAV follows its own motion model, path planning for UAVs is not feasible by design. Therefore, we first need to derive a preliminary clustering scheme based on the position information of the UAVs.

During the clustering phase, the satellite first collects the position information of all UAVs. Based on the position information of the UAVs, the position-based K-Means algorithm is used to divide the UAVs into \( c_n \) clusters, each with a CH and several CMs. An important factor affecting the effectiveness of the clustering algorithm is whether the optimal value of the number of clusters \( c_n \) can be determined. The number of clusters \( c_n \) affects the size of the clusters and the number of decisions and transmissions in the network. If the size of the clusters is large, the number of transmissions required for information exchange from CM UAVs to CH UAV will be very high, thereby affecting network performance. Conversely, if the scale of the clusters is too small, the number of clusters and CH UAVs will increase, and at this time, the number of decision-making agents increases, making decision conflicts more likely to occur, affecting the convergence performance of the network. Therefore, it is necessary to first determine the optimal value of \( c_n \), which is the minimum number of clusters that makes the maximum value of the CH coverage probability reach a preset threshold.

The CH coverage probability, that is, the probability that a UAV is covered by a CH, can be calculated based on the Euclidean distance between the position of the UAV and the position of the CH, as well as the communication radius of the CH \cite{bhandari2020mobility}. The closer the distance between the UAV and the CH, the greater this probability, and the more likely the UAV is to join the cluster of this CH. Specifically, it can be modeled using the logistic function \cite{reed2002application}, which is smooth and continuous, aligning with the characteristics of real-world communication systems. So the probability that UAV \( i \) is covered by CH \( l \) , denoted as \( P_{il} \), is expressed as:
\begin{equation}
	P_{il} = \frac{1}{1 + e^{-\zeta (d_{il} - R)}}
	\label{pi},
\end{equation}
where, \( d_{il} \) is the Euclidean distance between UAV \( i \) and CH \( l \), \( R \) is the communication radius of the UAV, and \( \zeta \) is a tuning parameter. The position information of CH \( l \) can be obtained based on the distribution of positions of each CH at the end of the previous clustering period \( T_{cls} \).

The cluster size, that is, the expected number of UAVs in a cluster, indicates on average how many CM UAVs are covered by this CH in a cluster. Therefore, it is calculated based on the CH coverage probability:
\begin{equation}
	E[N_l] = \sum_{i=1}^N P_{il},
\end{equation}
where, \( E[N_l] \) is the size of cluster \( l \), \( N \) is the total number of UAVs, and \( P_{il} \) is the probability that UAV \( i \) is covered by CH \( l \). Theoretically, due to the randomness of UAV positions, when UAVs are uniformly distributed, the cluster sizes of each cluster will be consistent. At this time, \( \kappa \) represents the number of UAVs per unit area. Since the communication radius of the UAV is a fixed value, the expected number of UAVs covered by the CH can be represented by the product of the density \( \kappa \) of UAVs and the area of the coverage area of the CH. Given the above assumption of uniform distribution, the theoretical value of the cluster size is:
\begin{equation}
	E[N_l] = \sum_{i=1}^N P_{il} = \kappa \pi R^2.
\end{equation}

The maximum value of the CH coverage probability, that is, the maximum probability that a UAV is covered by any CH, indicates that the closer the UAV is to all CHs, the greater this probability, and the more likely the UAV is to join any cluster. Obviously, it can be calculated based on the CH coverage probability as:
\begin{equation}
	P_{max} = 1 - \prod_{l=1}^{c_n} (1 - P_{il}),
	\label{pmax}
\end{equation}
where, \( P_{max} \) is the maximum value of the CH coverage probability, \( c_n \) is the number of clusters, and \( P_{il} \) is the probability that UAV \( i \) is covered by CH \( l \).

The larger the number of clusters \( c_n \), that is, the more CHs, the greater the maximum probability \( P_{max} \) that a UAV is covered by any CH, that is, the more likely it is to join any cluster. However, when the number of clusters \( c_n \) exceeds a critical value and becomes too large, the growth rate of the maximum probability \( P_{max} \) that a UAV is covered by any CH will also tend to slow down, that is, increasing the number of CHs has small effect on improving the coverage rate. At the same time, the larger the number of clusters \( c_n \), it also means that the complexity of network decision-making is higher, and the efficiency and stability of the network are lower. Therefore, it is necessary to find a balance point where \( P_{max} \) reaches a preset threshold, and the number of clusters \( c_n \) reaches a minimum value. This balance point is the optimal value of the number of clusters. In summary, the optimal value of the number of clusters is the minimum number of clusters that makes the maximum value of the CH coverage probability reach a preset threshold. It can be solved using the Lagrange multiplier method \cite{bhandari2020mobility}, resulting in:
\begin{equation}
	{c_n}^* = \frac{\ln (1 - P_{max})}{\ln (1 - \kappa \pi R^2 / N)},
\end{equation}
where, \( {c_n}^* \) is the optimal value of the number of clusters. Once the optimal number of clusters is determined, the preliminary clustering scheme can be determined according to the K-Means algorithm. The objective function of the position-based K-Means algorithm is to minimize the sum of the position distances of UAVs within a cluster.
\begin{equation}
	J=\sum_{i=1}^N d(\text{loc}_i, \text{loc}_{c_i}...),
\end{equation}
where \( N \) is the total number of UAVs, \( loc_i \) is the position of the \( i^{th} \) UAV, \( loc_{c_i} \) is the position of the cluster center to which the \( i^{th} \) UAV belongs, and \( d() \) is the position distance function.

At this point, the preliminary clustering scheme has been determined. The satellite broadcasts the clustering scheme to all UAVs until after \( T_{cls} \), when the clustering scheme is recalculated in the next clustering phase.

\subsubsection{Cluster Maintenance Algorithm}
During the cluster maintenance phase, the goal is to manage topological changes and membership variations due to UAV mobility. This includes two main scenarios: UAVs joining or leaving clusters and re-electing CHs. The process is distributed, with each UAV tracking its own location and that of the CH, while the CH maintains cluster structure information.

The CH, responsible for task scheduling, collects status information (location, motion state, and task queue) from the UAVs and makes scheduling decisions. It then broadcasts timestamped messages with its location.

CMs receive broadcasts from multiple CHs, compare their locations, and calculate a coverage probability. The higher the probability, the closer the UAV is to the CH, making it more suitable to join that cluster. The UAV selects the CH with the highest probability and sends a message to join. If no message is received, the UAV enters an isolated state.

Each CH regularly checks its list of CMs. If it discovers that members have left or joined, it updates its cluster structure. At each time slot, the CH calculates the central position of the cluster based on the location distribution information of the UAVs within the cluster and determines whether it is the UAV closest to the cluster center. Considering the stability of the clustering results, if the CH is not the cluster center for consecutive \( t_{ele} \) time slots, the CH is replaced with the latest cluster center. The original CH sends a replacement message to the new CH UAV and broadcasts a CH update message to the CMs, thereby achieving the re-election of the CH.

\begin{figure}[!t]
	\centering
	\includegraphics[width=0.48\textwidth]{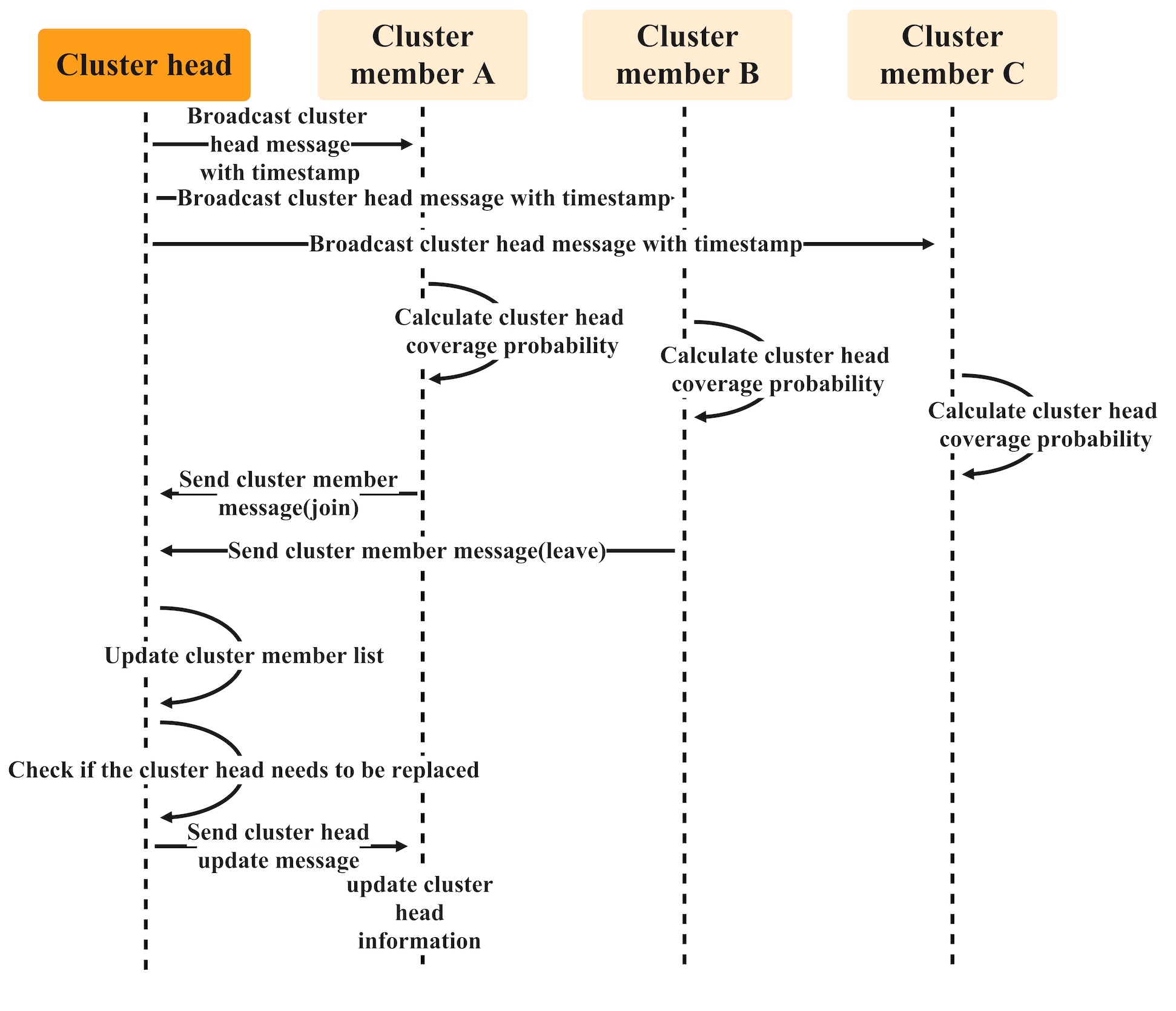}
	\caption{Schematic of the cluster maintenance algorithm.}
	\label{fencu}
\end{figure}

The overall interaction of the cluster maintenance algorithm is shown in Figure \ref{fencu}.

\subsubsection{Overall Process of the Dynamic UAV Clustering Algorithm}

The KMDUC algorithm is a location-based dynamic UAV clustering algorithm designed to adapt to the high mobility and self-organizing characteristics of the integrated air-space-ground network. The KMDUC algorithm is divided into two stages: the clustering stage and the cluster maintenance stage. The clustering stage, coordinated by satellites, uses the location-based K-Means algorithm based on the UAVs' location information to determine the number of clusters and CHs, as well as the initial assignment of CMs. The cluster maintenance stage, under the autonomous coordination and communication of UAVs, dynamically adjusts the structure of the cluster and the selection of the CH according to the dynamic changes of the UAVs, ensuring the stability and efficiency of the cluster. The following is a summary of the KMDUC algorithm process, with the overall process shown in Algorithm \ref{kmduc}.

\begin{algorithm}[t]
	\caption{KMDUC Algorithm}
	\label{kmduc}
	\Input{Set of UAVs $\mathcal{U}$, clustering period $T_{cls}$, CH replacement threshold $t_{ele}$\;}
	\Output{Clustering result for UAVs\;}
	Satellites initialize CHs, members, and stability counter\;
	\While{UAV set $\mathcal{U}$ is not empty}{
		\If{Clustering period reached $t \pmod{T_{cls}} = 0$}{
			Satellites collect UAV locations and calculate the optimal number of clusters ${c_n}^*$\;
			Satellites update and broadcast clustering results\;
		}
		\For{each CH $a_l \in A$}{
			$a_l$ collects member status and issues task scheduling\;
			$a_l$ broadcasts its location and checks if it is the center\;
			\If{$a_l$ is not the center}{
				\If{not the center for $t_{ele}$}{
					$a_l$ replaces the head and broadcasts update\;
				}
			}
		}
		\For{each member $U_i \in \mathcal{U}$}{
			$U_i$ sends status to head, executes task decisions\;
			$U_i$ receives head location and calculates coverage probability\;
			\eIf{$U_i$ finds greater coverage probability}{
				$U_i$ switches CH and sends update\;
			}{
				\If{$U_i$ receives no information from head}{
					$U_i$ becomes isolated\;
				}
			}
		}
	}
\end{algorithm}

\subsection{Multi-Agent Cooperative Offloading Algorithm}

Following the clustering process, the intelligent agents on each CH are responsible for the task offloading decisions within their cluster. In cooperative scenarios, we focus on how to enhance the cooperative interaction of agents to increase the overall system profit, while also aiming to improve training efficiency and convergence performance as much as possible.

To ensure the achievement of the optimization objectives stated in problem (\ref{problem})—maximizing the overall system profit—the reward function is set as follows:
\begin{equation}
	r(s,a)=\max_{\boldsymbol{\alpha}_{k,j}} \sum_{j=1}^{M} \sum_{k=1}^{L} \alpha_{k,j}G_{k},
\end{equation}
where, each agent's reward function represents the overall task profit of the system, thereby maximizing social welfare. The shared reward function ensures that the group of agents has a common optimization goal, promoting coordination and cooperation among agents, improving collective performance, and reducing variance during the learning process, as agents can learn from each other's experiences and actions.

CM UAVs periodically transmit their state information, including location and motion parameters, to the CH UAV via Line-of-Sight (LoS) wireless communication \cite{9922666}. When making offloading decisions, the CH UAV considers the location, motion characteristics, and task properties of the CM UAVs being decided upon. In cooperative scenarios, since the decisions for all task-collecting CM UAVs are made by the CH UAV, there is no competition for resources within the cluster. The CH UAV makes judgments based on the information of the entire cluster, thus specific resource allocation is no longer needed. The local observation space no longer includes information on the resource allocation vector but maintains the vector of all computational resources within the cluster coverage denoted as $\mathbf{R^*}$. In the design of the action space, the offloading priority of tasks is no longer independent per UAV but is shared within the entire cluster.

The local observation space $o_t$ includes four parts: computational resource information, motion information of the UAVs pending decision, the transmission rate vector of the UAVs pending decision, and current task feature information.

\begin{equation}
	\label{state}
	o_t=( \mathbf{R^*},\mathbf{M},\mathbf{Tr}, \mathbf{Task}).
\end{equation}

The resource vector $\mathbf{R^*}$ represents the computational resource information within the cluster coverage, including all computational resources of ground stations within the cluster coverage and the evenly distributed resources of satellites. Due to limited coverage, the computational resources of ground stations at the cluster center can be considered exclusively used by that cluster. On the other hand, the computational resources at the cluster edges are shared by all clusters, fully utilizing the complete cooperative relationship between clusters. The motion information $\mathbf{M} = (loc, \theta, v)$ includes the dynamic characteristics of the UAVs pending decision, such as their location, direction of movement, and speed. The transmission rate vector $\mathbf{Tr} = (tr^0, tr^1, \ldots, tr^m)$ captures the communication characteristics between the UAVs pending decision and each BS and satellite. Task requirements $\mathbf{Task}$ include the size of the task, computational complexity, and latency requirements.

The action space $a_t$ represents the offloading decision made by the CH agent for the $k$th computational task, consisting of two parts:
\begin{equation}
    \label{action}
	a_t = \{\alpha^*_k, \eta^*_k\} \quad \text{where} \quad \alpha_k \in \mathcal{D} \quad \text{and} \quad \eta_k \in [0, 1],
\end{equation}
where $\alpha^*_k$ indicates the target device to which the task $k$ will be offloaded. The second part of the action space is a continuous variable $\eta^*_k \in [0, 1]$, representing the computational priority of the task $k$ on the selected target device, with all computational tasks being processed on the target device in order according to $\eta^*_k$.

To better implement the cooperative mode of multi-agents, this paper combines the MADDPG algorithm to achieve centralized training and distributed execution in the integrated air-space-ground network. In this mode, the dispersed CH agents train a centralized Critic network based on the observations and actions of all agents, but act using their own distributed Actor network based on their local observations \cite{lowe2017multi}.

The algorithmic design of cooperative MADDPG consists of two main components: the Actor network for each agent and a Critic network. The Actor network is a deterministic policy that outputs actions based on the agent's observations, while the Critic network models a state-action value function, estimating the expected return for each agent given the global joint observations and actions of all agents. The key difference between MADDPG and single-agent DDPG is that the Critic network takes the joint global observations and actions as input, while the Actor network only takes individual observations as input.
\begin{figure}[!t]
	\centering
	\includegraphics[width=0.48\textwidth]{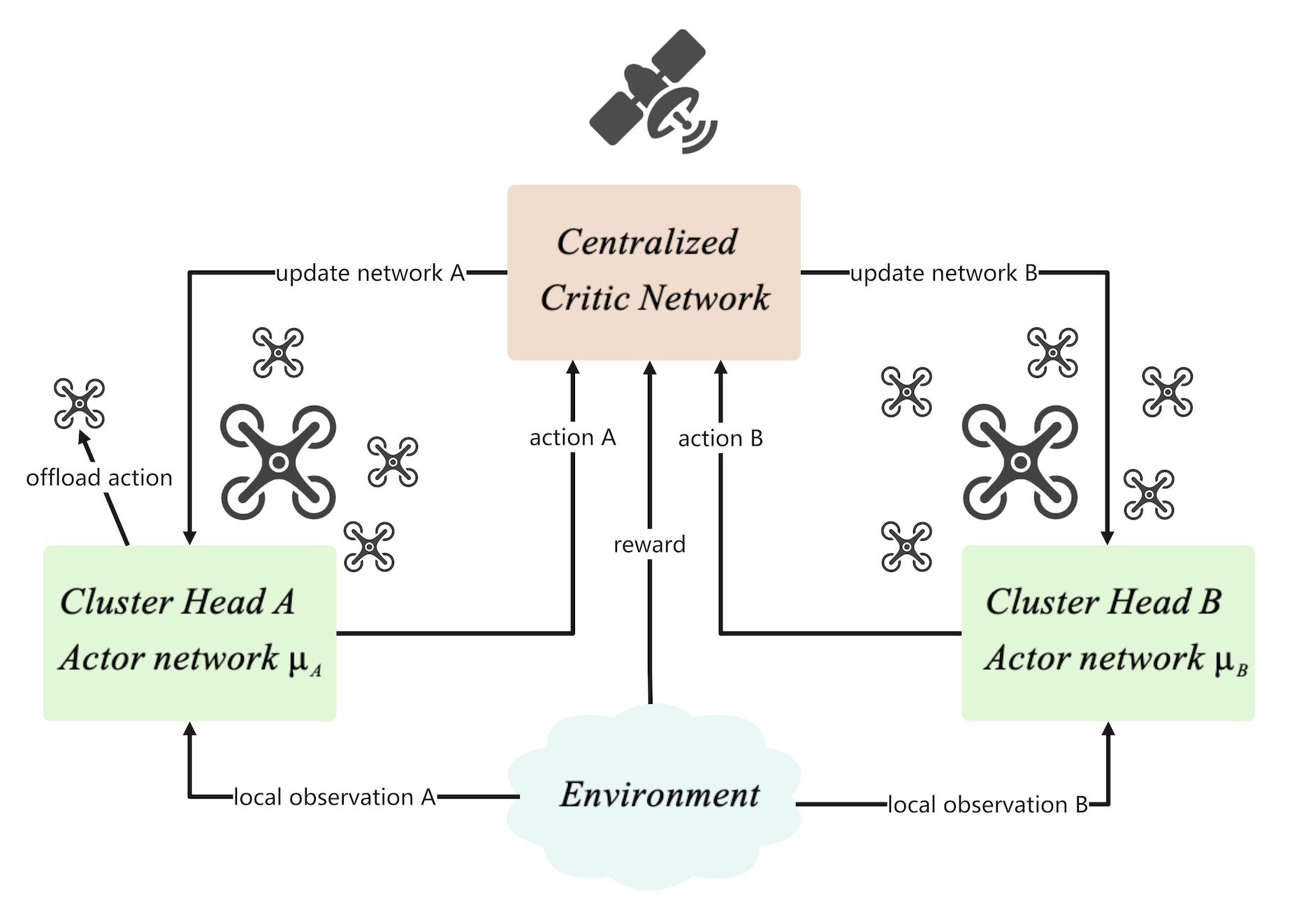}
	\caption{Centralized training and distributed execution cooperative framework.}
	\label{maddpg}
\end{figure}

This paper deploys the centralized training process on satellites, which collect global observation and action information to train the Global Critic network while placing the Actor network on each lead UAV for task offloading decisions. The centralized training and distributed execution cooperative MADDPG work mode is shown in Figure \ref{maddpg}.

In centralized training, each agent can access the joint observations and actions of all agents, as well as the global reward function. This allows agents to learn a centralized Critic network that can estimate the Q value for each agent based on the joint observations and information of all agents. The Critic network updates by minimizing the mean squared error between the target Q value and the current Q value:
\begin{equation}
	L_i(\theta_i^Q) = \mathbb{E}_{(o, a, r, o') \sim \mathcal{D}} \left[ \left( Q_i(o, a | \theta_i^Q) - y_i \right)^2 \right],
\end{equation}
where, $o = (o^1, \dots, o^N)$ is the joint observation of all agents, $a = (a^1, \dots, a^N)$ is the joint action of all agents, and $y_i = r_i + \gamma Q_i'(o', a' | \theta_i^{Q'})$ is the target Q value, calculated using the target Critic network $Q_i'$ and target Actor network $\mu_i'$, where $a'^i = \mu_i'(o'^i | \theta_i^{\mu'})$.

The gradient of the loss function for the Critic network parameters is:
\begin{align}
\nabla_{\theta_i^Q} L_i(\theta_i^Q) = 
& \, 2 \mathbb{E}_{(o, a, r, o') \sim \mathcal{D}} \Big[ \left( Q_i(o, a | \theta_i^Q) - y_i \right) \nonumber \\
& \quad \times \nabla_{\theta_i^Q} Q_i(o, a | \theta_i^Q) \Big].
\end{align}

Critic network parameters are updated by applying the stochastic gradient descent method:
\begin{equation}
	\theta_i^Q \leftarrow \theta_i^Q - \alpha \nabla_{\theta_i^Q} L_i(\theta_i^Q).
	\label{critic}
\end{equation}

In distributed execution, each agent can only use its own local observations and selects an action using its own distributed Actor network. The Actor network is a deterministic policy that maps the agent's observation to an action. The Actor network updates by applying the policy gradient method, maximizing the expected Q value of the Critic network as follows:
\begin{align}
\nabla_{\theta_i^\mu} J(\theta_i^\mu) = 
& \, \mathbb{E}_{o \sim \mathcal{D}} \Big[ \nabla_{\theta_i^\mu} \mu_i(o_i | \theta_i^\mu) \nonumber \\
& \quad \cdot \nabla_{a_i} Q_i(o, a | \theta_i^Q) \big|_{a_i = \mu_i(o_i)} \Big],
\end{align}
where, $o_i$ is the individual observation of agent $i$, and $\mu_i(o_i)$ is the individual action of agent $i$. Actor network parameters are updated by applying the stochastic gradient ascent method:
\begin{equation}
	\theta_i^\mu \leftarrow \theta_i^\mu + \alpha \nabla_{\theta_i^\mu} J(\theta_i^\mu).
	\label{Actor}
\end{equation}

For the target networks, parameters are updated through soft updates.
\begin{align} 
	\theta_i^{\mu'} \leftarrow \tau \theta_i^\mu + (1 - \tau) \theta_i^{\mu'}.
    \label{soft}
\end{align}

The soft update method ensures a stable learning process by gradually blending the parameters of the target networks with those of the learned networks. This approach avoids drastic changes in the policy, which could lead to divergence or instability in the learning process. The parameter controls the rate at which the target networks are updated, with a smaller $tau$ resulting in slower updates and a more stable learning process.

In summary, the cooperative MADDPG framework facilitates the centralized training of a global Critic network that leverages the full state and action space, while enabling distributed execution where each agent acts based on its local observations and a distributed Actor network. This approach allows for the efficient coordination of multiple agents in complex environments, such as the integrated air-space-ground network, where communication constraints and dynamic scenarios are prevalent. The cooperative nature of the algorithm ensures that all agents work towards a common goal, maximizing the overall system profits and promoting a cohesive multi-agent collaboration.

The paradigm of centralized training and distributed execution enables the algorithm to cope with partial observability, non-stationarity, and communication challenges in multi-agent environments. This is because agents can learn from global information during training and act independently during execution. Moreover, agents can optimize the global reward function while maintaining their individual strategies, and they do not need to communicate or coordinate with each other during execution, reducing communication overhead and the risk of communication failures. By utilizing centralized training and distributed execution, the algorithm fully leverages the advantages of satellite coverage and broadcasting for efficient collection and dissemination of parameters. It also adapts well to the highly dynamic nature of the environment, using UAVs for flexible task scheduling and fully adapting to the multi-dimensional heterogeneity of the integrated air-space-ground network.

The cooperative offloading algorithm is as shown in Algorithm \ref{alg:coop}. In each time slot, each CM UAV collects its local observations and sends them to the CH UAV. The CH UAV uses its Actor network to make offloading decision actions, issues them to the corresponding CM UAVs, and stores them in the experience pool. The satellite regularly collects local observations and actions from the CH UAVs according to its \(T_{update}\) update cycle, forms joint behaviors and actions to be stored in the experience pool, updates the Critic and Actor networks according to Equations (\ref{critic}) and (\ref{Actor}), and then distributes the updated Actor network parameters to each CH UAV.

\begin{algorithm}[t]
	\caption{Multi-Agent Cooperative Offloading Algorithm}
	\label{alg:coop}
	\Input{Set of $c_{num}$ CH agents with Actor networks, global Critic network, and experience pool\;}
	\Output{Task offloading strategy\;}
	Initialize global Critic network $Q$, target network $Q'$, random weights $\theta^Q$, $\theta^{Q'}$\;
	Initialize each agent's Actor network $\mu_i$, target network $\mu_i'$, random weights $\theta_i^\mu$, $\theta_i^{\mu'}$\;
	Initialize experience pool $D$ and reward function\;
	\For{each training cycle $T_{update}$}{
		\For{each time slot}{
			\For{each CM UAV}{
				Collect local observation, and send it to CH UAV $i$\;
			}
			CH UAV $i$ selects offloading decision $a_i = \mu_i(o_i | \theta_i^\mu)$\;
			CH issues action, member UAV executes it\;
			CH observes reward $r_i$ and next local observation $o_i'$\;
			Store transition $(o_i, a_i, r_i, o_i')$ in experience pool $D$\;
		}
		Sample batch $B$ from experience pool $D$\;
		Update global Critic network $Q$ using equation (\ref{critic})\;
		Update each agent's Actor network using equation (\ref{Actor})\;
		Soft update global Critic and each agent's target network using equation (\ref{soft})\;
		Send updated Actor parameters $\theta_i^\mu$ to each CH UAV $i$\;
	}
\end{algorithm}

\subsection{Overall Process of Cluster-Based Cooperative Offloading Algorithm for Multi-Agent Systems}

In summary, the CMADDPG algorithm is a cluster-based cooperative task scheduling approach for multi-agent systems, consisting of two components: the KMDUC UAV clustering algorithm and the multi-agent cooperative offloading algorithm. Its goal is to achieve coordinated control and intelligent scheduling of UAVs in an integrated air-space-ground network, enhancing system efficiency and profits. The algorithm is outlined in Algorithm \ref{alg:cmaddpg}.

The CMADDPG algorithm first uses satellites for coordination and the K-Means algorithm to cluster UAVs needing task scheduling, determining the number of clusters, CHs, and initial member allocations. Next, it adapts dynamically by adjusting cluster structures and CH selections to respond to UAV mobility, ensuring stability and efficiency. The algorithm then deploys intelligent agents on CH UAVs to make offloading decisions based on their Actor networks. These decisions include task offloading targets, and computation priorities, and are stored in the experience pool. Finally, the algorithm leverages satellite coverage to periodically collect data from the experience pool and Actor network parameters, training the Critic and Actor networks. The updated Actor network parameters are distributed to the CH UAVs for more efficient offloading decisions.

\begin{algorithm}[t]
	\caption{CMADDPG Algorithm}
	\label{alg:cmaddpg}
	\Input{Integrated air-space-ground network environment\;}
	\Output{Task offloading decisions\;}
	Initialize environment and neural networks\;
	\While{within the operation cycle}{
		\If{reaching clustering period $T_{cls}$}{
			Satellites compute optimal clusters ${c_n}^*$ and broadcast the clustering results\;
		}
		\For{each time slot}{
			\For{each UAV}{
				Join/leave cluster, collect local observation, send to CH UAV $i$\;
			}
			CH UAV $i$ selects offloading decision and stores transition\;
			\For{each CH UAV}{
				Update and broadcast CH\;
			}
		}
		Satellites collect experience pool and network parameters from each CH UAV\;
		Satellites update global Critic and CH Actor networks\;
		Satellites distribute updated network parameters\;
	}
\end{algorithm}

\subsection{Complexity analysis}
% 我们将算法分为分簇和卸载两部分分析算法复杂度：
% \subsubsection{KMDUC Algorithm}
% 根据算法1，第8行到第13行复杂度为O(c_n^*)，第14到第21行复杂度为O(N_u)，假设while循环最多L轮结束，因此复杂度为O(L(c_n^*+N_u))=O(L(N_u))
We divide the algorithm into two parts: clustering and offloading, and analyze the algorithm complexity separately:

\subsubsection{KMDUC Algorithm}  
According to Algorithm \ref{kmduc}, the complexity from Lines 8 to 13 is \( O(c_n^*) \), and the complexity from Lines 14 to 21 is \( O(N_u) \). Assuming the \texttt{while} loop terminates after at most \( L \) iterations, the overall complexity is:  $O(L (c_n^* + N_u)) = O(L N_u)$

\subsubsection{Multi-Agent Cooperative Offloading Algorithm}
% 神经网络的复杂度主要取决于前向传播和反向传播时每层见的矩阵运算。设observation的维度根据公式（20）设为dim_o,action的维度根据公式（21）设为dim_a,设actor network有n_a层隐藏层，每层隐藏层的维度为e_a，则actor network的复杂度为O_{actor}=O(dim_o\times e_a +(n_a-1)\times e_a^2+e_a\times dim_a).设critic network有n_c层隐藏层，每层隐藏层的维度为e_c，则critic network的复杂度为O_{critic}=O((dim_o+dim_a)\times e_c +(n_c-1)\times e_c^2+e_c\times 1).
The complexity of a neural network mainly depends on the matrix operations in each layer during forward propagation and backward propagation. Let the dimension of the observation, according to Equation (\ref{state}), be denoted as $\text{dim}_o$, and the dimension of the action, according to Equation (\ref{action}), be denoted as $\text{dim}_a$. Assume the actor network has $n_a$ hidden layers, and the dimension of each hidden layer is $e_a$. Then, the complexity of the actor network is: $O_{\text{actor}} = O(\text{dim}_o \times e_a + (n_a - 1) \times e_a^2 + e_a \times \text{dim}_a).$ Assume the critic network has $n_c$ hidden layers, and the dimension of each hidden layer is $e_c$. Then, the complexity of the critic network is: $O_{\text{critic}} = O((\text{dim}_o + \text{dim}_a) \times e_c + (n_c - 1) \times e_c^2 + e_c \times 1).$

% 由于CMADDPG算法先将大规模无人机分簇，进一步通过簇首进行决策，算法复杂度为各无人机独立决策的\frac{c_n^*}{N_u}，又由于在大规模场景中c_n^*<<N_u，因此CMADDPG算法更加快捷高效。

Since the CMADDPG algorithm first clusters the large-scale UAVs and further delegates decision-making to the CHs, the algorithm complexity is reduced to $\frac{c_n^*}{N_u}$ of that when each UAV makes independent decisions. Furthermore, in large-scale scenarios where $c_n^* \ll N_u$, the CMADDPG algorithm is significantly faster and more efficient.

\section{Performance Evaluation}

In this section, we validate the effectiveness of the CMADDPG algorithm through numerical simulation experiments. We first introduce the integrated air-space-ground network simulation environment and the basic settings of the algorithm, then present the performance metrics, and finally conduct a series of simulation experiments to verify the algorithm's performance compared to benchmark algorithms.

\subsection{Experimental Parameters}

We conduct simulation experiments within a larger area. This is due to the consideration that resource competition is often influenced by the communication range of UAVs and BSs, which are concentrated in smaller areas, while UAV cooperation still exists extensively within larger coverage areas. This section conducts simulations within a square area with a side length of 5 km, which is four times the area of the competitive scenario simulation setup, deploying 25 task processing BSs with computational capabilities. The computational power of these BSs varies between 20 GHz and 40 GHz. In the task scheduling scenario, 40 UAVs are deployed, with their movement speeds following a normal distribution. The arrival of computing tasks follows a Poisson distribution with a given arrival rate, while task size, computational load, and latency requirements follow a uniform distribution. Table \ref{hezuocanshu} provides key parameters used in the simulation under the cooperative scenario.

\begin{table}[!ht]
	\centering
	\caption{Simulation Parameters for Cooperative Scenario}
	\label{hezuocanshu}
	\begin{tabularx}{\columnwidth}{lX}
		\hline
		\textbf{Parameter} & \textbf{Value} \\
		\hline
		Area Size & $5$ km $\times$ $5$ km \\
		Number of BSs & $25$ \\
		Number of UAVs & $40$ \\
		Computational Power Range & $20$ GHz - $40$ GHz \\
		UAV Movement Speed & Mean: $10$ m/s; Standard Deviation: $2$ m/s \\
		Task Arrival Distribution & $\mu$: $25$ tasks/s \\
		Task Size Distribution & 10 Mbits - 90 Mbits \\
		Computational Load Distribution &  1000 M Cycles - 3000 M Cycles \\
		Tolerable Latency Distribution & 0 ms - 200 ms \\
		Batch Size & $64$ \\
		Actor network Learning Rate & $0.01$ \\
		Critic network Learning Rate & $0.001$ \\
		Experience Pool Size & $100000$ \\
		\hline
	\end{tabularx}
\end{table}

In terms of the neural networks, both the Actor and Critic networks contain two hidden layers. The Actor network has 128 neurons in both layers, while the Critic network has 256 neurons in one layer and 128 neurons in another, with different learning rates for both, as shown in Table \ref{hezuocanshu}. The training still employed the Adam optimizer and $tanh$ activation function. For exploration noise, the experiment used the Ornstein-Uhlenbeck process to generate noise \cite{lowe2017multi}, with a mean of 0, a standard deviation of 0.3, and a volatility value of 0.15.

The comparative methods chosen for this experiment are the naive MADDPG algorithm used in the literature \cite{seid2021multi} and the task offloading method of Multi-Agent Actor-Critic (MAAC) used in the literature \cite{zhu2021learning}. Compared to the CMADDPG algorithm of this paper, the MADDPG algorithm lacks a clustering process, so each UAV independently makes offloading decisions without following the command of the CH, leading to an increase in the number of offloading agents in the network. In the MAAC algorithm, each agent has an independent Critic network, which is different from the CMADDPG and MADDPG algorithms.

\subsection{Results}

\begin{figure}[!t]
	\centering
	\includegraphics[width=0.48\textwidth]{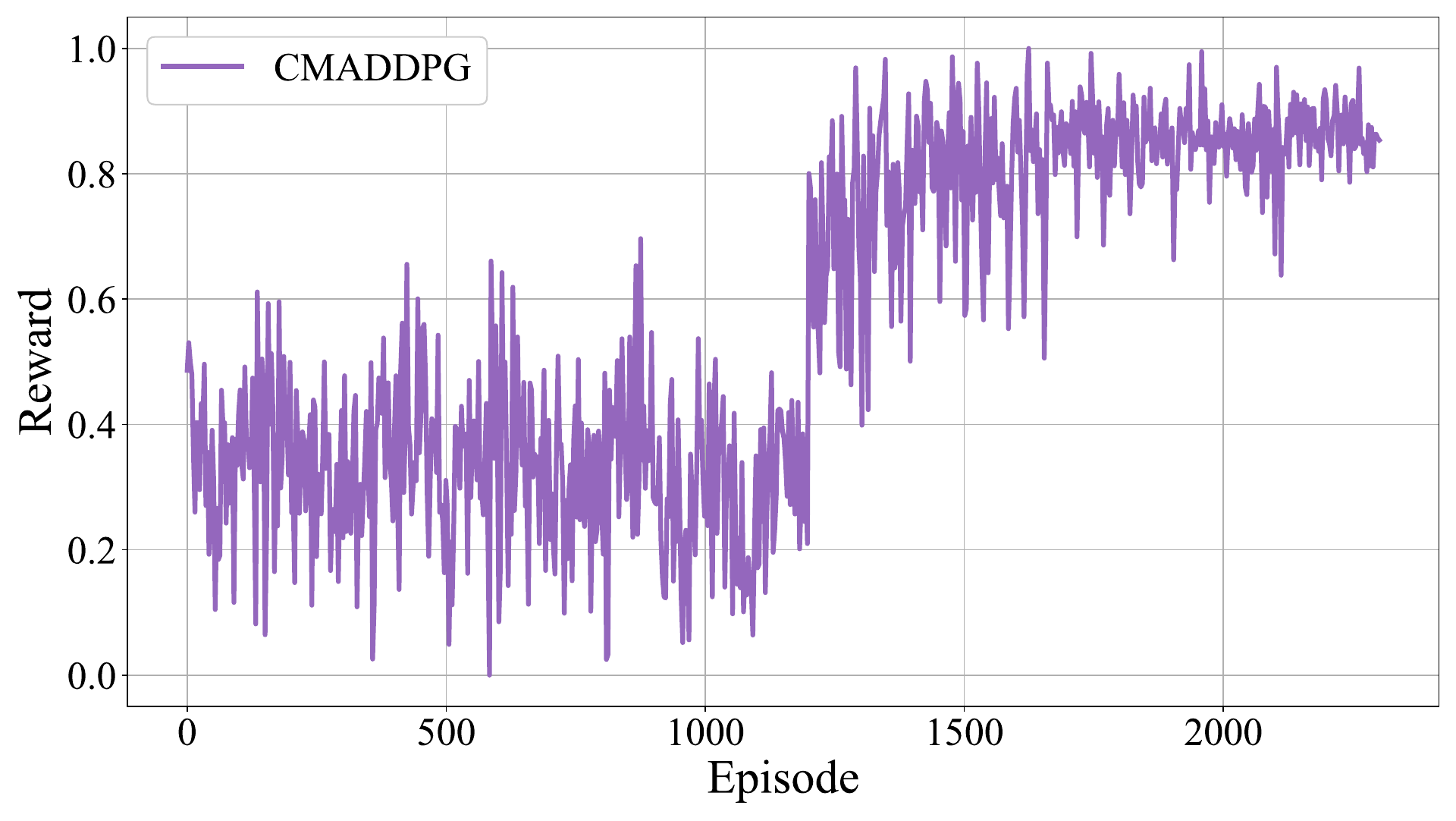}
	\caption{Reward during training cycles.}
	\label{shoulll}
\end{figure}
Figure \ref{shoulll} shows the change in the reward of CMADDPG with the training cycles, reflecting the convergence performance. It can be seen that in the early stages of training, the agents are in the stage of exploring the environment. Around 1100 training cycles, the curve experiences a steep increase, indicating the rapid convergence performance of the multi-agent model after completing exploration, reflecting the performance advantage of model convergence under the cooperative mode of centralized training and distributed execution, allowing the model to quickly adapt to complex cooperative scenarios.
\begin{figure}[!t]
	\centering
	\includegraphics[width=0.48\textwidth]{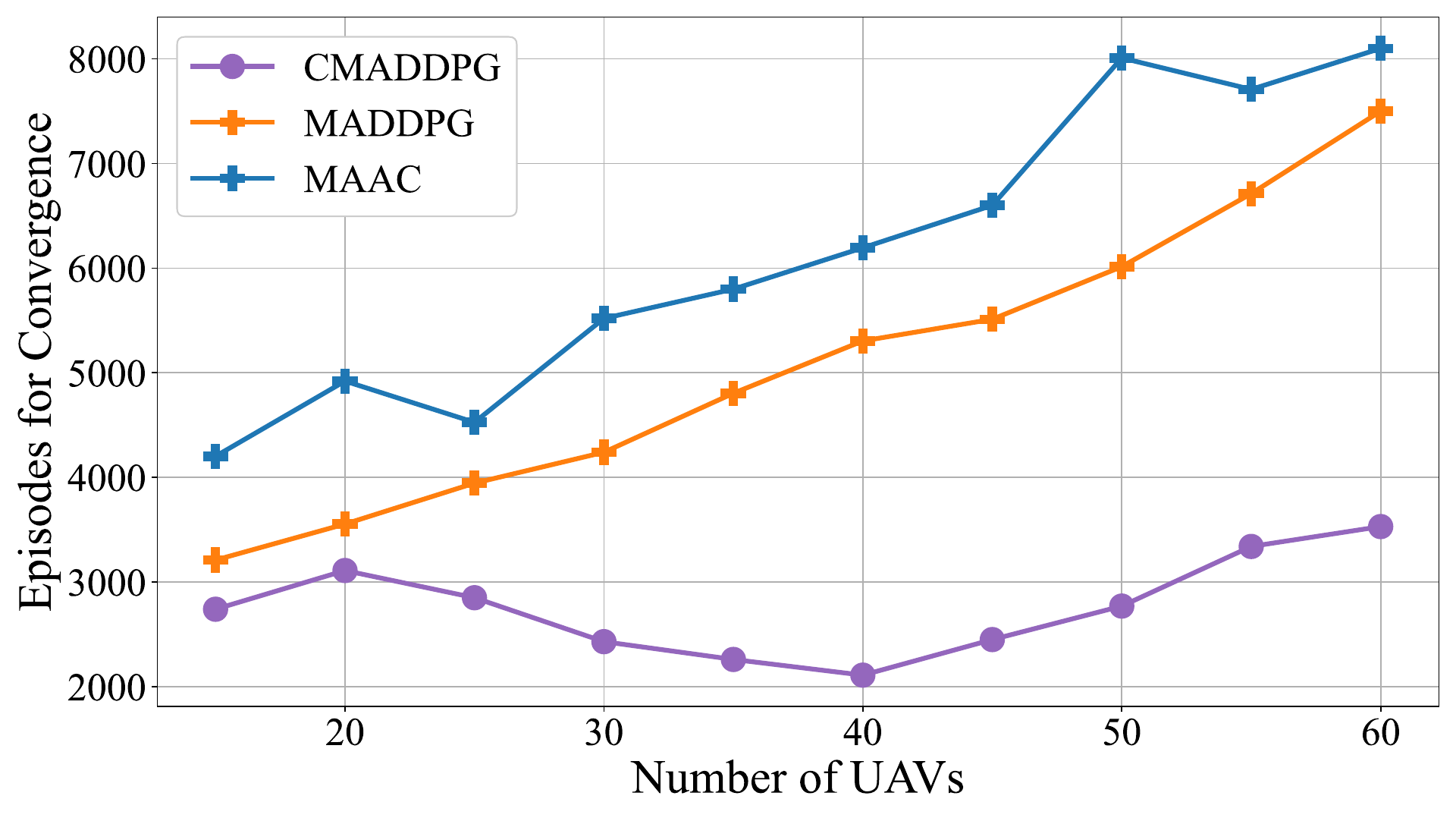}
	\caption{Episodes for convergence as UAV count increases.}
	\label{zhouqi}
\end{figure}

Figure \ref{zhouqi} shows the change in the number of cycles required for convergence of each algorithm as the number of UAVs increases. It is clear from the figure that the training duration required for the CMADDPG algorithm is significantly lower than that of the MADDPG and MAAC algorithms, benefiting from the clustering algorithm that deploys agents only on the CH UAVs, thus reducing the number of UAVs that need to make decisions. In addition, the number of cycles required for convergence of the CMADDPG algorithm first decreases and then increases with the number of UAVs. This is because when the number of UAVs is small, they are distributed sparsely over a wide area, making it difficult to form clusters, leaving many isolated UAVs. As the number of UAVs increases, isolated UAVs can gradually be connected into clusters, reducing the number of agents required for offloading decisions, thus improving convergence performance. For the MADDPG and MAAC algorithms, the time required for convergence increases with the number of agents, even showing an exponential growth trend. The convergence performance of the MAAC algorithm is significantly longer because it does not adopt the cooperative mode of centralized training and distributed execution, and the algorithm itself lacks robustness.

\begin{figure}[!t]
	\centering
	\includegraphics[width=0.48\textwidth]{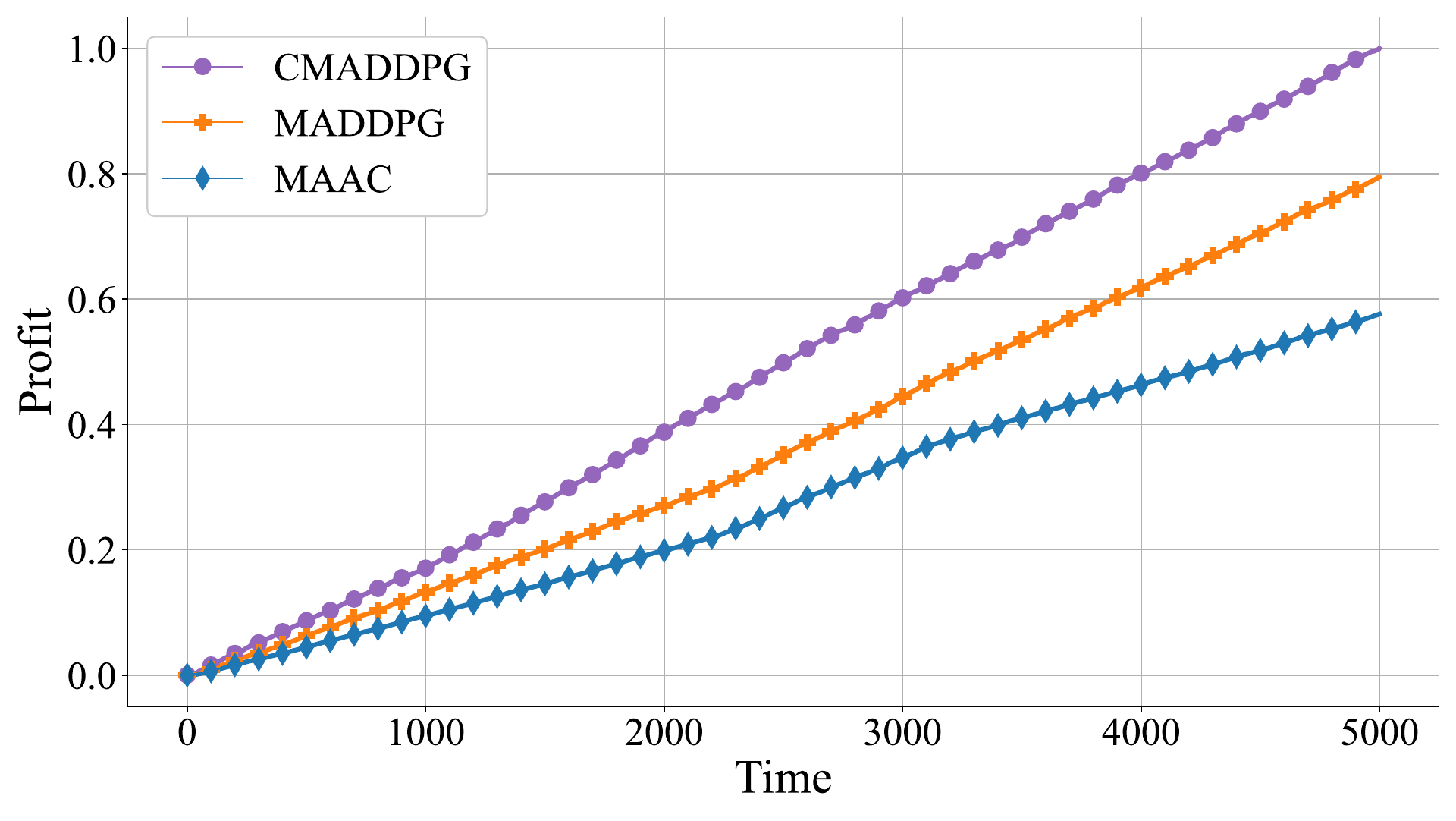}
	\caption{Overall system profit over time.}
	\label{shijian}
\end{figure}
Figure \ref{shijian} shows the change in the overall system profit over time. Compared to the competitive scenario, the fluctuation of the overall system profit has significantly increased, which is caused by the increased dynamism due to the increase in the number of UAVs within a larger range. Evidently, the CMADDPG algorithm exhibits the best overall performance, achieving a 25\% improvement in system profit compared to MADDPG and a 42.86\% improvement compared to MAAC. This is attributed to the clustering algorithm, where the CH can better utilize the information of all members within the cluster for decision-making and training, resulting in a more stable curve.

Although MADDPG has utilized the sample data of other agents in terms of training efficiency, there is still a lack of communication and cooperation among agents during the decision-making process, resulting in higher volatility of the curve. MAAC, due to the insufficiency of the algorithm itself, shows disadvantages in all aspects.
\begin{figure}[!t]
	\centering
	\includegraphics[width=0.48\textwidth]{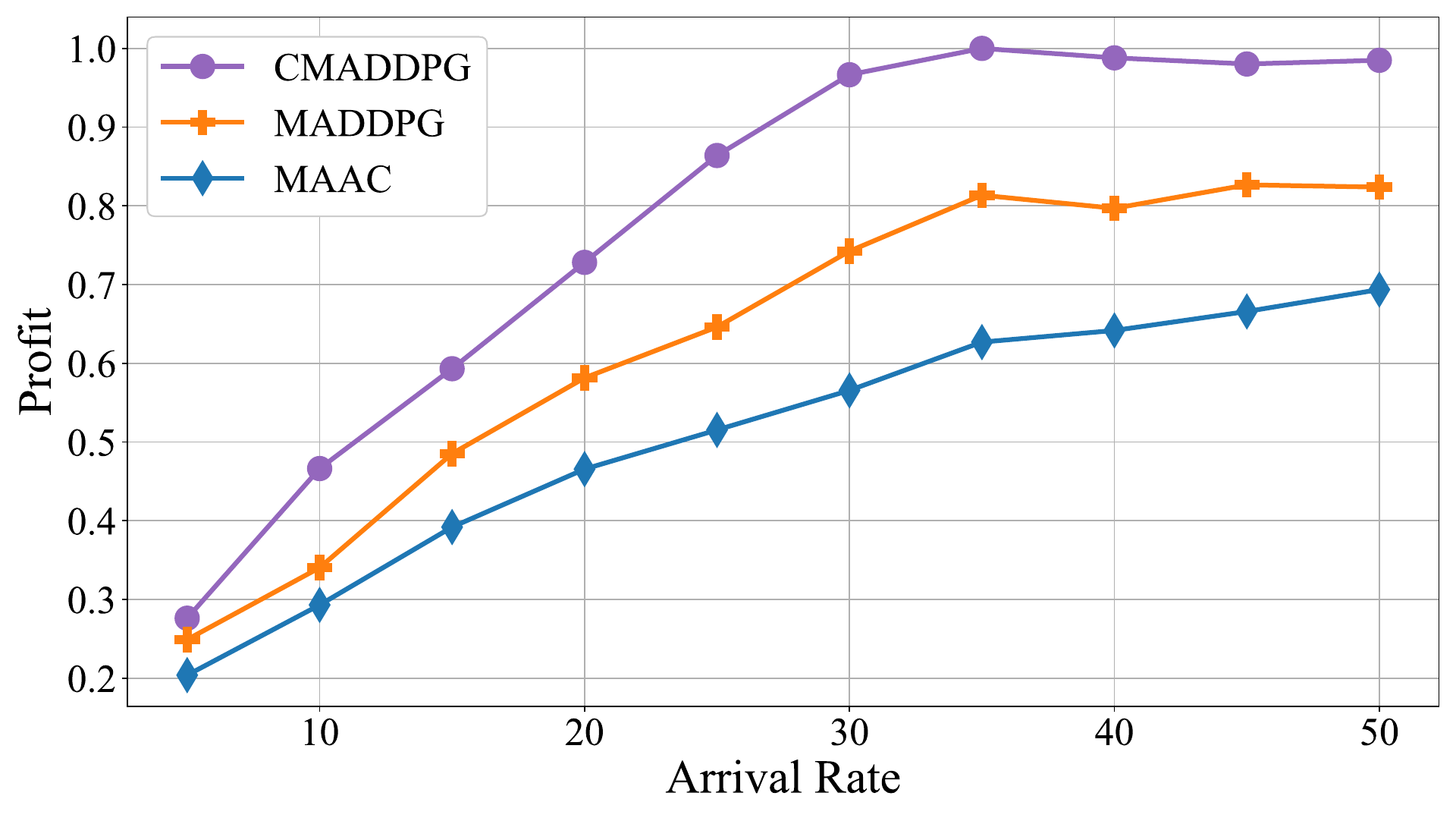}
	\caption{System profit under varying loads determined by task arrival rates.}
	\label{hezuodaoda}
\end{figure}

Figure \ref{hezuodaoda} describes the change in system profit with the change in system load, which is determined by the task arrival rate. The higher the task arrival rate, the heavier the system load, indicating that the system needs to handle more tasks. The CMADDPG algorithm shows excellent performance under both low and high loads, converging earlier, indicating that it utilizes system resources more fully earlier on. The benchmarked algorithm MADDPG shows similar load convergence performance, reaching a high point in profits at a task arrival rate close to that of the CMADDPG algorithm, but under the same load conditions, its scheduling of tasks is not as good as the CMADDPG algorithm, with decisions based only on local observation information leading to insufficient profits. The MAAC algorithm maintains growth in profits over a longer load interval, but the profits are always insufficient, indicating poor task scheduling decision performance.

\begin{figure}[!t]
	\centering
	\includegraphics[width=0.48\textwidth]{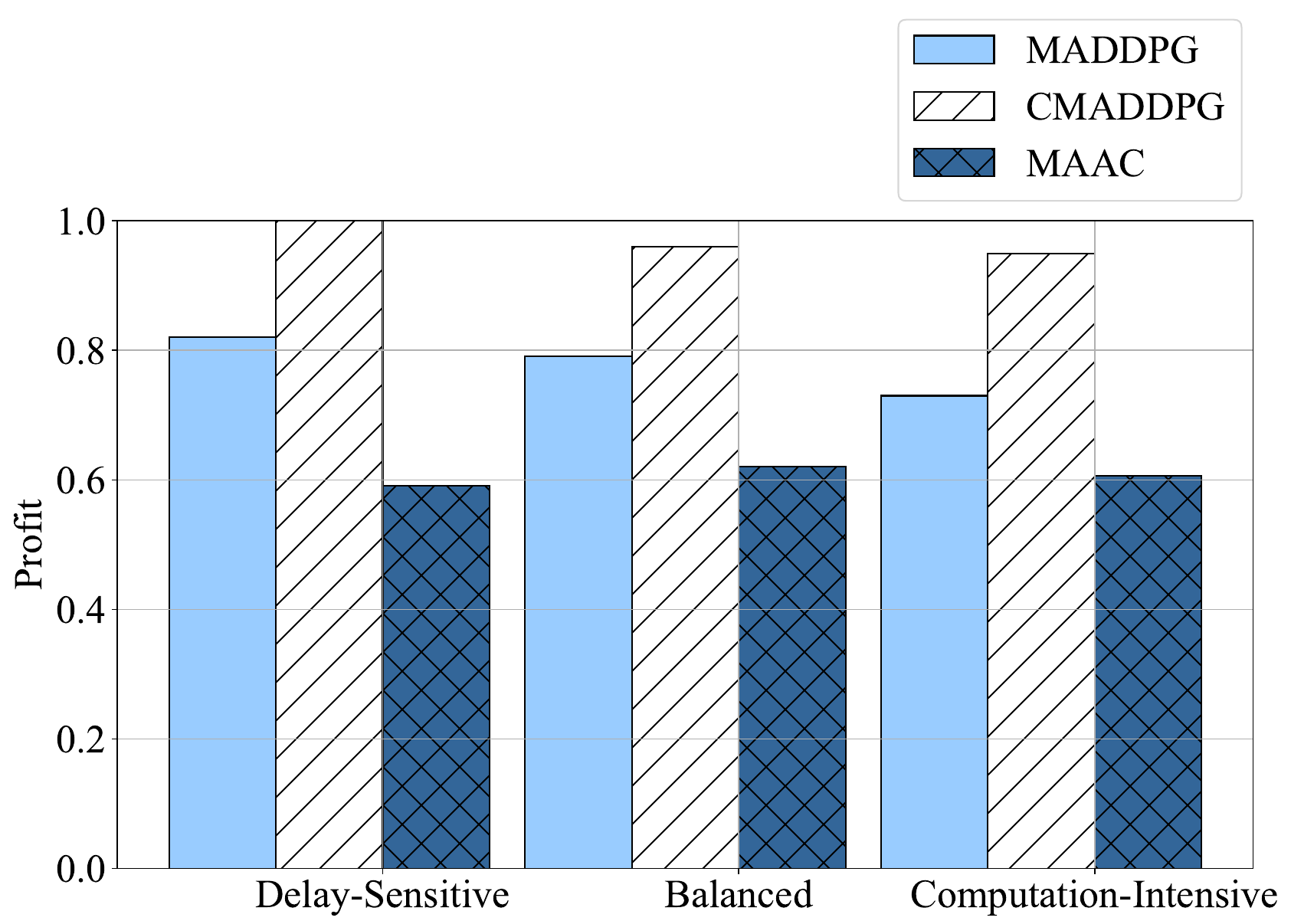}
	\caption{Normalized system profit in three scenarios.}
	\label{shouyifenbu}
\end{figure}

Figure \ref{shouyifenbu} shows the normalized system profit under three different scenarios: delay-sensitive, computation-intensive, and balanced distribution. In the experiment, in the delay-sensitive scenario, the probability of generating tasks with latency requirements in the $0-20$ms range is increased, with the proportion of delay-intensive tasks in the scenario increased to $30 \% $, while increasing the weight parameter $\lambda^{\rho}$ for latency in the task profits function. For the computation-intensive scenario, the probability of generating larger computational tasks is increased, while reducing the weight parameter $\lambda^{\rho}$ for latency in the task profits function. In the balanced distribution scenario, the experimental settings remain unchanged.

As can be seen under different task distributions, the CMADDPG algorithm maintains the optimal profits, demonstrating the performance guarantee of the CMADDPG algorithm under more stringent scenarios. In the delay-sensitive scenario, the profits of the CMADDPG algorithm are higher than in the balanced distribution, indicating that the CMADDPG algorithm has fully prioritized the scheduling of delay-sensitive tasks and obtained higher profits from them, reflecting the intelligence and adaptability of the CMADDPG algorithm. However, in the computation-intensive scenario, the system's profits decrease, which is due to the reduction of $\lambda^{\rho}$ leading to a decrease in the profits obtained per unit of task computation.

The MADDPG algorithm maintains a trend similar to that of the CMADDPG algorithm but with a certain performance gap. The system profit of the MAAC algorithm do not vary much across different task distribution scenarios, indicating its lack of adaptability to the environment.

\begin{figure}[!t]
\centering
\includegraphics[width=0.48\textwidth]{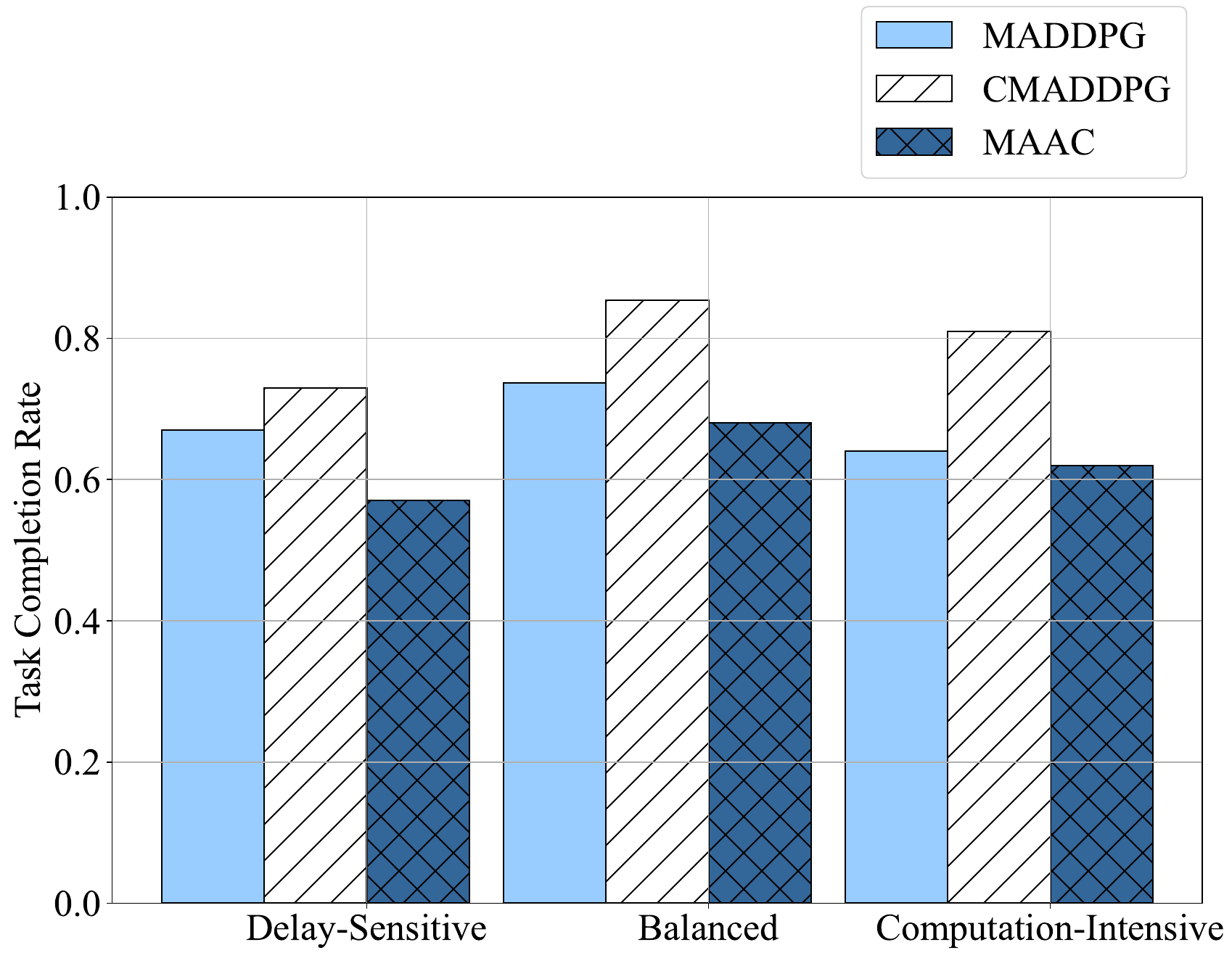}
\caption{Task completion rate in three scenarios.}
\label{fenbu}
\end{figure}
Figure \ref{fenbu} shows the task completion rate under three different scenarios: delay-sensitive, computation-intensive, and balanced distribution. It can be observed that under different task distributions, the CMADDPG algorithm maintains the optimal task completion rate. Combined with Figure \ref{shouyifenbu}, the CMADDPG algorithm, with a lower task completion rate in the delay-sensitive scenario, achieves higher system profit, indicating that the CMADDPG algorithm can prioritize the scheduling of delay-sensitive tasks under resource-limited stringent scenarios, demonstrating the robustness of its scheduling decisions.

\section{Conclusion}
This paper has addressed the task scheduling problem in cooperative SAGIN scenarios and has proposed the CMADDPG algorithm. The algorithm first applies the K-Means algorithm for centralized clustering of UAVs and has performed distributed cluster maintenance within each cluster, introducing the KMDUC algorithm for a hybrid distributed-centralized control approach. Building on this, we have proposed a cooperative offloading algorithm based on MADDPG, leveraging satellite-based information dissemination and parameter sharing. In this algorithm, each cluster head acts as an agent making offloading decisions in a distributed manner, while a centralized evaluation platform optimizes these decisions, enabling centralized training and distributed execution of the offloading strategy. Simulation results have demonstrated favorable and rapid convergence characteristics, validating a significant improvement in system profit, with a 25\% increase in system profit compared to benchmark methods. The proposed scheme has effectively utilized the collaborative capabilities of satellites, UAVs, and base stations in SAGIN, combining the advantages of centralized and distributed control to reduce communication costs, resolve decision conflicts, and improve network efficiency, showcasing the immense potential of intelligent collaborative networks in future complex communication environments.

\bibliographystyle{IEEEtran}
\bibliography{ref.bib} 

\end{document}